\pdfoutput=1
\documentclass[11pt,twoside,a4paper,cmspaper,final,collab]{cms-tdr}
\def\svnVersion{31aefc0}\def\svnDate{2026/09/30}\def\cmsCernNoTag{CERN-EP-2026-050}\def\cmsCernDate{\today}\def\cmsMessage{Published in Physical Review Letters as \href{http://dx.doi.org/10.1103/ll3f-l5xv}{\doi{10.1103/ll3f-l5xv}.}}
\begin{document}\cmsNoteHeader{HIN-24-013}

\newlength\cmsFigWidth
\ifthenelse{\boolean{cms@external}}{\setlength\cmsFigWidth{0.49\textwidth}}{\setlength\cmsFigWidth{0.65\textwidth}}
\ifthenelse{\boolean{cms@external}}{
\newcommand{\cmsLeft}{upper\xspace}
\newcommand{\cmsRight}{lower\xspace}
}{
\newcommand{\cmsLeft}{left\xspace}
\newcommand{\cmsRight}{right\xspace}
}

\newcommand{\PbIon}{\ensuremath{\mathrm{Pb}}\xspace}
\newcommand{\PbPbIon}{\ensuremath{\mathrm{PbPb}}\xspace}
\newcommand{\MuMu}{\ensuremath{\PGmp\PGmm}\xspace}
\newcommand{\fD}{\ensuremath{f_{\mathrm{D}}}\xspace}
\newcommand{\fI}{\ensuremath{f_{\mathrm{I}}}\xspace}
\newcommand{\RgPb}{\ensuremath{R^{\mathrm{Pb}}_{\Pg}}\xspace}
\newcommand{\WgN}{\ensuremath{W_{\PGg\mathrm{N}}}\xspace}
\newcommand{\nGama}{\ensuremath{n_{\PGg}}\xspace}
\newcommand{\STARLIGHT} {\textsc{STARlight}\xspace}

\cmsNoteHeader{HIN-24-013}

\title{Observation of  nuclear suppression in coherent \texorpdfstring{\PgUa}{Upsilon(1S)} photoproduction off heavy nuclei at the LHC}

\date{\today}

\abstract{
The first measurement of coherent \PgUa meson photoproduction off  heavy nuclei is performed using ultraperipheral lead-lead collisions collected by the CMS experiment at a nucleon-nucleon center-of-mass energy of 5.02\TeV. The nuclear gluonic structure is probed at a nucleon momentum fraction of order $x{\sim}10^{-3}$, determined by the kinematics of the process. Owing to the large \PgUa mass, the measurement reaches  the highest scale accessible so far through coherent vector-meson photoproduction, $\mu^2 = 22.4\GeV^2$, where nonlinear quantum chromodynamics effects are expected to be minimal. 
In the \PgUa rapidity range $\abs{y}<1$, the ratio of the measured photoproduction cross section to a baseline model prediction that neglects nuclear effects is $S_{\PgUa}=0.25\pm0.06\stat\pm0.02\syst$, thereby demonstrating nuclear suppression in this process. Expressed in terms of  a nuclear gluon suppression factor, the result yields $\RgPb(x\approx10^{-3},\mu^2=22.4\GeV^2)=0.55\pm0.12\stat\pm0.02\syst$.
The measured $S_{\PgUa}$ is only slightly larger than the values previously reported for coherent \PGf photoproduction, despite the probed $\mu^2$ differing by approximately two orders of magnitude. 
}

\hypersetup{
pdfauthor={CMS Collaboration},
pdftitle={Observation of nuclear suppression in coherent Upsilon(1S) photoproduction off heavy nuclei at the LHC},
pdfsubject={CMS},
pdfkeywords={CMS, UPCs, saturation, nPDFs}
}

\maketitle

At nucleon momentum  fractions of order $x\lesssim10^{-3}$, the structure of free nucleons and of nucleons bound in nuclei becomes increasingly gluon dominated~\cite{H1:2015ubc,Frankfurt:2005mc}. As $x$ decreases further, the gluon density rises rapidly because of gluon splitting~\cite{Bjorken:1968dy}.
At sufficiently small $x$, nonlinear quantum chromodynamics (QCD) effects, such as gluon recombination, are expected to moderate this growth and eventually lead to a novel regime known as gluon saturation~\cite{Iancu:2003xm,Gelis:2010nm}.
Heavy nuclei, which have densities significantly higher than those of free nucleons, offer an ideal environment to search for gluon saturation.
Investigating QCD dynamics at extreme gluon densities and probing gluon saturation have motivated extensive experimental efforts in heavy ion collisions~\cite{Klein:2020fmr,Bertulani:2005ru,CMS:2024krd} and constitute a major physics motivation for the future Electron Ion Collider~\cite{Accardi:2012qut}.

In ultraperipheral collisions (UPCs) of heavy nuclei, the ions interact at impact parameters greater than the sum of their radii, so that strong interactions are negligible and electromagnetic interactions dominate, providing a clean environment for studying photon-photon and photonuclear interactions~\cite{Bertulani:2005ru,Baltz:2007kq,Klein:2020fmr}. Among these processes, the photoproduction of heavy vector mesons is particularly significant. At leading order (LO), a quasireal photon from the projectile ion fluctuates into a quark-antiquark dipole, which subsequently interacts with the target nucleus through the exchange of two gluons in a color-singlet configuration, forming a vector meson. This process can serve as a sensitive probe of the gluonic structure of the target nucleus~\cite{Ryskin:1992ui,Brodsky:1994kf,Accardi:2012qut}. In UPCs,  photoproduction  occurs through either coherent or incoherent processes. Coherent photoproduction, where a photon interacts with the entire nucleus, probes the average gluon distribution, producing vector mesons with low transverse momentum ($\pt<0.2\GeV$). Conversely, incoherent photoproduction probes local gluon distributions and their fluctuations, resulting in vector mesons with substantially higher \pt~\cite{Guzey:2013jaa,Guzey:2016piu,Mantysaari:2020axf}.

Coherent photoproduction probes the target nucleus at a momentum-transfer scale largely set by the square of the vector meson mass, \ie, $\mu^2{\sim}M^2/4$. Measuring different vector mesons therefore enables complementary access to the nuclear gluon structure and to possible nonlinear QCD effects, each with distinct sensitivity~\cite{Guzey:2016piu,Mantysaari:2017dwh,Bendova:2020hbb}.
Coherent photoproduction of $\rho$, \PJGy, and \PGyP{2S}  mesons in UPCs has been extensively studied at the BNL RHIC and CERN LHC energies~\cite{Afanasiev:2009hy,Adamczyk:2017vfu,ALICE:2013wjo,Khachatryan:2016qhq,ALICE:2021gpt,ALICE:2020ugp,LHCb:2021bfl,CMS:2023snh,ALICE:2023jgu,STAR:2023nos,ATLAS:2025aav,CMS:2025oxg}.
These measurements have revealed significant nuclear modifications to the production cross sections, offering key tests for theoretical models of gluon distributions in nuclei.
Recent studies by the CMS~\cite{CMS:2023snh} and ALICE~\cite{ALICE:2023jgu} experiments have provided the first measurements of the dependence of coherent \PJGy photoproduction on  the photon-nucleon center-of-mass energy (\WgN).
The results reveal significantly greater nuclear suppression at higher \WgN (or equivalently lower $x$) compared to the baseline predictions by the impulse approximation (IA)~\cite{Guzey:2013xba,Guzey:2018tlk,Kryshen:2023bxy}, which neglects nuclear effects by treating the nucleus as a collection of free nucleons.
The CMS Collaboration has further extended these studies with a \WgN-dependent measurement of incoherent \PJGy photoproduction~\cite{CMS:2025oxg}, which shows an even stronger suppression toward lower $x$ compared to the coherent case. Remarkably, the ratio of incoherent to coherent photoproduction remains constant over a wide \WgN range of $\approx$40--400\GeV (corresponding to $x=5.9\times10^{-3}$ to $6.5\times10^{-5}$).
The recent CMS measurement of coherent \PGf meson photoproduction in lead-lead (\PbPbIon) UPCs~\cite{CMS:2025lsm} finds a nuclear suppression factor of 0.18--0.20 at $x\sim10^{-4}$ and $\mu^{2}\approx0.26\GeV^{2}$
, lower than the corresponding values obtained for coherent \PJGy photoproduction by the ALICE~\cite{ALICE:2021gpt} and ATLAS experiments~\cite{ATLAS:2025aav}, thereby providing evidence for the mass dependence of nuclear modifications in vector-meson photoproduction~\cite{Forshaw:2004vv,Kowalski:2006hc,Frankfurt:2005mc}.
Despite these experimental advances, significant challenges remain for state-of-the-art theoretical models that include gluon saturation~\cite{Mantysaari:2017dwh,Bendova:2020hbb,Lappi:2010dd,Cepila:2016uku,Mantysaari:2023xcu,Mantysaari:2022sux} or nuclear shadowing effects~\cite{Guzey:2016piu,Mantysaari:2017dwh,Guzey:2018tlk,Kryshen:2023bxy,Henkels:2023plt,Jenkovszky:2022qnc,Jenkovszky:2023jrh}, underscoring the need for continued experimental and theoretical advances.

Coherent \PgU photoproduction occurs at a $\mu^2{\sim}9$ times that of \PJGy and ${\sim}90$ times that of \PGf, accessing a distinct kinematic regime.
Next-to-LO (NLO) perturbative QCD (pQCD) calculations~\cite{Eskola:2022vpi} show that both quark and gluon contributions are important for \PJGy, challenging the assumption that at LO the cross section scales trivially with the square of the nuclear gluon parton distribution function (PDF). In contrast, quark contributions are predicted to be negligible for coherent \PgU production~\cite{Eskola:2023oos}, suggesting that it may serve as a more precise probe of the nuclear gluon structure of heavy nuclei. Although exclusive \PgU photoproduction off protons was previously studied in proton-lead UPCs at $\sqrtsNN=5.02\TeV$~\cite{CMS:2018bbk}, coherent \PgU photoproduction off heavy nuclei in UPCs remained elusive because of its extremely low production rate and the large background from photon-photon processes.

This Letter presents the first measurement of coherent \PgUa photoproduction off a heavy nucleus in UPCs. The \PgUa signal is reconstructed via the \MuMu decay channel using data collected by the CMS detector in 2018 during \PbPbIon collisions at the LHC at a nucleon-nucleon center-of-mass energy of $\sqrtsNN=5.02\TeV$. The data sample corresponds to an integrated luminosity of $1.66\pm0.03$\nbinv~\cite{CMS:2025rzq}. Differential cross sections are presented  in two intervals of the \PgUa meson rapidity, $\abs{y}<1$ and $1<\abs{y}<2.4$. Additionally, the \PgUa production cross section per $\PGg\PbIon$ interaction is measured at midrapidity ($\abs{y}<0.2$), corresponding to $\WgN\approx200\GeV$ and probing gluons at $x{\sim}10^{-3}$. Tabulated results are provided in the HEPData record for this analysis~\cite{HEPData}.

The CMS apparatus~\cite{CMS:2008xjf} is a multipurpose, nearly hermetic detector, designed to trigger on~\cite{CMS:2020cmk,CMS:2016ngn} and identify electrons, muons, photons, and hadrons~\cite{CMS:2020uim,CMS:2018rym,CMS:2014pgm}.  A global ``particle-flow'' (PF) algorithm~\cite{CMS:2017yfk} aims to reconstruct all individual particles in an event, combining information provided by the all-silicon tracker and by the crystal electromagnetic and brass-scintillator hadron calorimeters, operating inside a 3.8\unit{T} superconducting solenoid, with data from the gas-ionization muon detectors interleaved with the layers of the flux-return yoke outside the solenoid. 
Forward hadron calorimeters (HFs)~\cite{Bayatian:2006jz}, made of steel and quartz fibers, extend the pseudorapidity ($\eta$) coverage provided by the barrel and end-cap detectors.

Events are selected online using a dedicated trigger that requires at least one muon candidate, with no explicit \pt requirement, in coincidence with a \PbPbIon bunch crossing~\cite{CMS:2016ngn}. At the trigger level, events with energy deposits above noise thresholds in either forward calorimeter are vetoed.
In the offline selection, candidate events are required to contain a primary vertex reconstructed from at least two tracks, following the standard CMS Run 2 heavy ion tracking procedure~\cite{CMS:2024krd}. The vertex is required to lie within 25\cm along the beam direction and 2\cm transversely from the detector center. To suppress hadronic interactions, the energy deposits in the positive- and negative-$\eta$ HF calorimeters must be below thresholds determined from empty bunch-crossing events.
Events are required to contain exactly two tracks within $\abs{\eta}<2.4$. After all other event selection requirements have been applied, the two track requirement removes less than 2\% of the events in the sample.
The CMS ``soft muon'' identification criteria~\cite{CMS:2012nsv}, which rely on combined information from the silicon tracker and muon detectors, are applied to select high-quality muons.
Each muon is required to have a $\pt>3.5\GeV$ to ensure high reconstruction efficiency.
At least one of the two muon candidates must match the muon that triggered the event.
The \PgU meson candidates are constructed by pairing oppositely charged muons with an invariant mass in the range $8.0<m_{\MuMu}<12.0\GeV$.
The combinatorial background is evaluated using same-sign muon pairs and found to be negligible after applying the event- and muon-level selection criteria.

\begin{figure}
    \centering
    \includegraphics[width=0.495\textwidth]{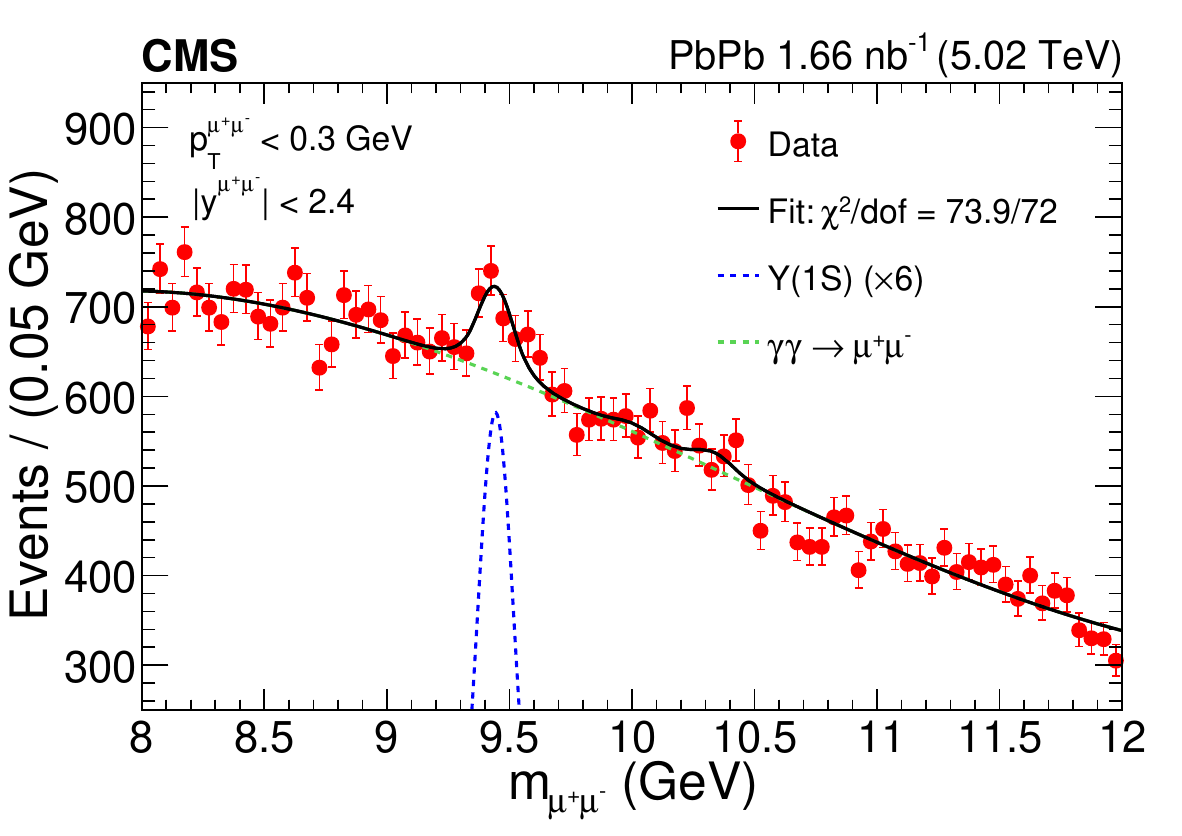}
    \includegraphics[width=0.495\textwidth]{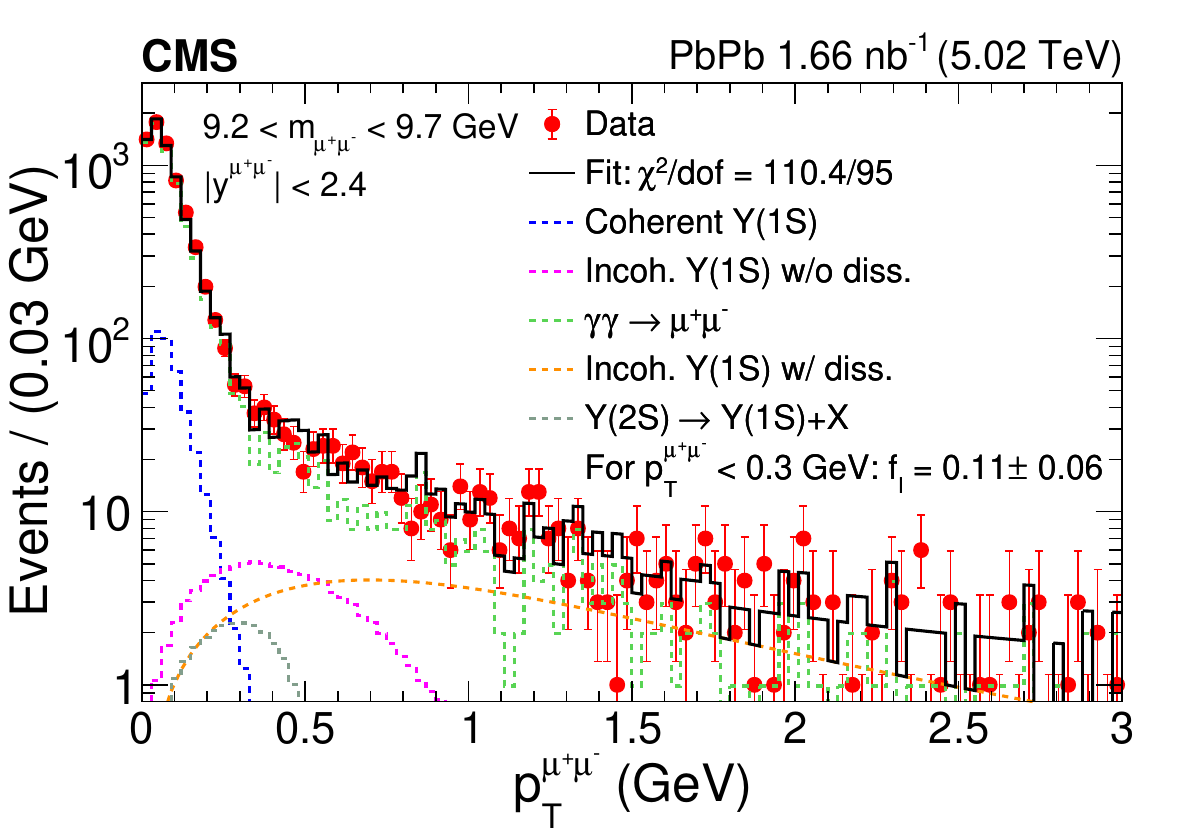}
    \caption{
        Invariant mass distribution of muon pairs with $\pt<0.3\GeV$ (\cmsLeft), where coherent production processes are dominant, and the transverse momentum distribution of muon pairs in the \PgUa mass window $9.2<m_{\MuMu}<9.7\GeV$ (\cmsRight).
        The total fit model and the individual fit components are shown as separate curves and are described in the text. The goodness of fit is quantified by $\chi^2/\text{dof}$, where dof denotes the number of degrees of freedom. In the $m_{\MuMu}$ fit, the \PgUa signal is scaled for visibility whereas the \PgUb and \PgUc components are not drawn.  
    }
    \label{fig:massFits}
\end{figure}

The reconstructed dimuon sample contains contributions from both coherent and incoherent processes, as well as a dominant quantum electrodynamics (QED) background from the $\PGg\PGg\to\MuMu$ production. The coherent component includes direct \PgUa production and the \PgUa mesons from the feed-down decays of the \PgUb and \PgUc states.
The incoherent contribution arises primarily from \PgUa photoproduction in both elastic and dissociative incoherent interactions.
The raw yield of \PgUa mesons is extracted by fitting the  $m_{\MuMu}$ and dimuon $\pt$ distributions, following the methodology established in Refs.~\cite{ALICE:2019tqa,ALICE:2021gpt,CMS:2023snh}, as illustrated in Fig.~\ref{fig:massFits}.

The initial raw yield of coherent \PgUa mesons, $N_{\PgUa}$, is extracted from a fit to the $m_{\MuMu}$ distribution in the region $\pt<0.3\GeV$, where coherent production dominates. The \PgU signal shape is modeled using a double-sided Crystal Ball (CB) function~\cite{Oreglia:1981fx}, characterized by four parameters, two describing the low- and high-mass tails, and two corresponding to the mean ($\mu_{\PgU}$) and width ($\sigma_{\PgU}$) of the distribution.
The tail parameters and width are fixed to values obtained from Monte Carlo (MC) simulations generated with \STARLIGHT (v3.13)~\cite{Klein:2016yzr} and processed through the full CMS detector simulation based on \GEANTfour~\cite{GEANT4:2002zbu}. This choice is supported by the inclusive ($\abs{y}<2.4$) fit, where the tail parameters are consistent with the MC within uncertainties, while in $\abs{y}$ bins the available data are insufficient to constrain them.
The mean and width of the \PgUb and \PgUc states are constrained by $\mu_{\PGUP{nS}}=\mu_{\PgUa} M_{\PGUP{nS}}/M_{\PgUa}$ and $\sigma_{\PGUP{nS}}=\sigma_{\PgUa} M_{\PGUP{nS}}/M_{\PgUa}$, using their world-average mass ($M_{\PGUP{nS}}$) values~\cite{ParticleDataGroup:2024cfk}.
The dominant physics background, arising from the QED process $\PGg\PGg\to\MuMu$, is described by a third-order polynomial.

The extracted $N_{\PgUa}$ yield includes contributions from direct coherent and incoherent production, as well as feed-down processes.
The contamination from incoherent contributions in the region $\pt<0.3\GeV$ is estimated using a template fit to the dimuon \pt distributions within the \PgUa mass window $9.2<m_{\MuMu}<9.7\GeV$.
The \pt templates for coherent \PgUa, elastic incoherent \PgUa, and feed-down contributions are obtained from the MC simulations, while the QED background template is derived from the sidebands at $8.8<m_{\MuMu}<9.2\GeV$ and $10.5<m_{\MuMu}<10.6\GeV$.
The dissociative incoherent \PgUa component is modeled using the empirical parameterization developed by the H1~\cite{H1:2013okq}, and then employed by the ALICE~\cite{ALICE:2019tqa,ALICE:2021gpt,ALICE:2023jgu}, ATLAS~\cite{ATLAS:2025aav}, and CMS~\cite{CMS:2023snh,CMS:2025oxg} Collaborations for heavy vector-meson photoproduction, $\rd N/\rd\pt\propto\pt\,[1+(b_{\mathrm{pd}}/n_{\mathrm{pd}})\pt^{2}]^{-n_{\mathrm{pd}}}$, where $b_{\mathrm{pd}}$ and $n_{\mathrm{pd}}$ are treated as free parameters in this analysis.
The normalizations of the coherent and incoherent (elastic and dissociative) contributions are left as free parameters in the fit. The total \PgUa yield can therefore be determined independently from the \pt fit and is used only as a consistency check, giving a result compatible with the baseline invariant mass extraction.
  The normalization of the QED component is fixed to the yields determined from the invariant mass fits.
Owing to limited statistical precision for the excited states, the ratio of feed-down to direct coherent production (\fD) in the coherent-dominated \pt region cannot be extracted from data, unlike in previous charmonium measurements~\cite{ALICE:2019tqa,ALICE:2021gpt,ALICE:2023jgu,CMS:2023snh,CMS:2025oxg}.
Instead, it is estimated using MC simulations, with \STARLIGHT generating the \PgU states and \PYTHIA8 (v303) modeling the decays of the excited \PgU states to \PgUa. Including the full CMS detector simulation, the resulting \fD is approximately $6.4\pm0.1$\% and is used to constrain the feed-down contribution in the \pt fits. The systematic uncertainty in \fD is estimated by varying the branching fraction of the excited \PgUb state decaying into the ground \PgUa state within the world-average uncertainty~\cite{ParticleDataGroup:2024cfk}.  The ratio of incoherent to coherent contributions, \fI, as determined by the fitting procedure, is $8\pm3\,\%$ ($13\pm5\,\%$) at central (forward) rapidities.

The raw yield of coherent \PgUa photoproduction is obtained as
\begin{equation}
N_{\PgUa}^{\text{coh}}=\frac{N_{\PgUa}}{1+\fI+\fD},
\end{equation}
and the corresponding rapidity-differential cross section is given by
\begin{equation}
\frac{\rd\sigma^{\text{coh}}_{\PgUa}}{\rd y}=\frac{N^{\text{coh}}_{\PgUa}}{(\mathcal{A}\epsilon)_{\PgUa}\mathcal{B}_{\PgUa\to\MuMu{}}\mathcal{L}_{\text{int}} \Delta y},
\end{equation}
where $\mathcal{A}$ and $\epsilon$ denote the detector acceptance and total reconstruction efficiency, respectively, $\mathcal{L}_{\text{int}}$ is the integrated luminosity, and $\Delta y$ is the rapidity bin width. The $\mathcal{A}$ and $\epsilon$ values are evaluated using coherent $\PgUa\to\MuMu$ MC samples generated with \STARLIGHT and processed through the full CMS detector simulation. Residual data-to-simulation differences in muon reconstruction efficiencies are corrected using scale factors derived with the tag-and-probe method~\cite{CMS:2012nsv,CMS:2011aa}. 
The \PgUa\ mass peak in the coherent-dominated $\pt<0.3\GeV$ region has an observed local significance of more than five standard deviations, evaluated using the one-sided profile-likelihood-ratio method implemented in the CMS Higgs Combine tool~\cite{CMS:2024onh}. The excited-state yields and the QED background normalization and shape parameters are profiled in the likelihood. The impact of signal-extraction modeling is assessed by considering alternative signal and QED background functions, as well as variations of the CB tail parameters and mass resolution, as described below. While the combined contribution of the excited states is constrained by the fit, the relative populations of the \PgUb and \PgUc components cannot be reliably resolved with the available statistical precision and are therefore not individually interpreted.

Systematic uncertainties are evaluated by considering variations from the nominal result for each given source.
The relative uncertainties in the integrated luminosity ($\mathcal{L}_{\text{int}}$) and the branching fraction ($\mathcal{B}_{\PgUa\to\MuMu}$) are 1.7\%~\cite{CMS:2025rzq} and 2\%~\cite{ParticleDataGroup:2024cfk}, respectively.
The choice of the HF calorimeter thresholds results in an uncertainty of 6\% (2\%), while the efficiency corrections from the tag-and-probe method introduce an uncertainty of 2\% (3\%), in the central (forward) rapidity interval. The uncertainties related to signal yield extraction are estimated by varying the signal and background models in the fit, as follows: (i) using a CB function combined with a Gaussian; (ii) modeling the QED continuum with a fourth-order polynomial; (iii) leaving the CB tail parameters free; (iv) varying the \PgUa width to values obtained from peripheral \PbPbIon collisions; and (v) changing the sideband invariant mass ranges for the QED \pt template determination. The model dependence of the \pt templates was further assessed by propagating bin-by-bin fluctuations of the data-based QED template and by varying the \PbIon nuclear radius and skin thickness~\cite{CMS:2023snh} used in the coherent \PgUa template. These variations have a negligible effect on the extracted coherent yield. The combined uncertainty in the signal extraction is 8\% (19\%) in the central (forward) rapidity interval. Adding the individual sources in quadrature yields a total systematic uncertainty of 10\% (20\%).

\begin{figure}
    \centering
    \includegraphics[width=\cmsFigWidth]{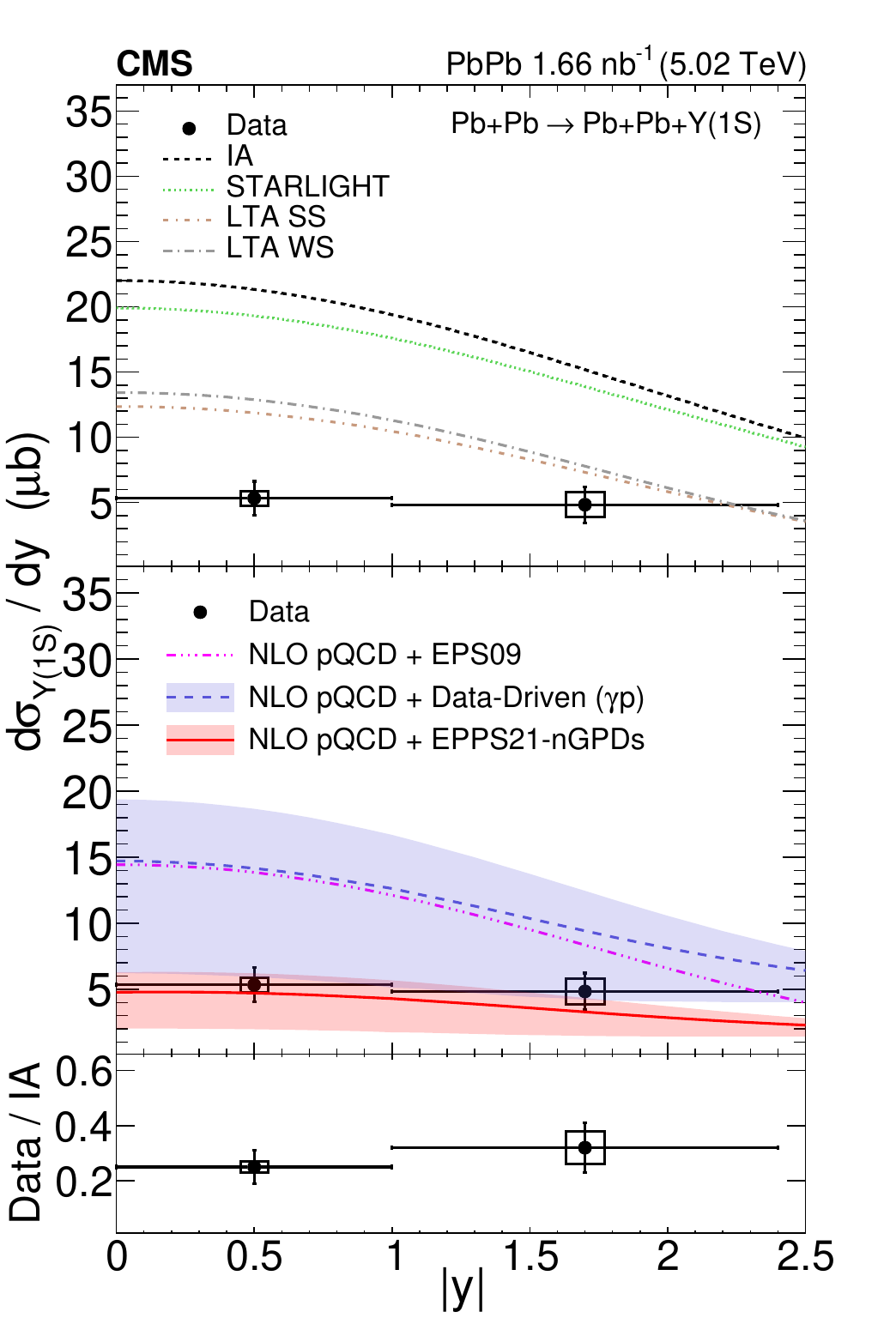}
    \caption{
        Differential cross section for coherent \PgUa production in the central $\abs{y}<1.0$ and forward  $1.0<\abs{y}<2.4$ intervals.
        The vertical error bars and boxes represent the statistical and systematic uncertainties, respectively. The results are compared with theoretical predictions from the impulse approximation (IA)~\cite{Guzey:2013jaa, Klein:2016yzr}, \STARLIGHT~\cite{Klein:2016yzr}, and LTA model with two nuclear shadowing scenarios~\cite{Guzey:2016piu} in the top panel,  and with three NLO pQCD calculations using the nuclear PDFs EPS09 or EPPS21, or  a data-driven scaling constrained by $\PGg\Pp$ data~\cite{Guzey:2016piu, Eskola:2023oos} in the middle panel. The bottom panel shows the ratio of the data to the IA prediction. The uncertainty in the ratio is experimental only and does not include any uncertainty in the IA baseline.
    }
    \label{fig:xSec}
\end{figure}

Figure~\ref{fig:xSec} shows the results in the central
interval of $\abs{y}<1.0$ and the forward interval of
$1.0 < \abs{y} < 2.4$. This represents the first measurement of coherent \PgUa photoproduction off a heavy nucleus. The results in the two $\abs{y}$ intervals are consistent within uncertainties, indicating no significant $y$ dependence. 

The data are compared with several theoretical models in Fig.~\ref{fig:xSec}. The IA model~\cite{Guzey:2013jaa,Klein:2016yzr} serves as the baseline corresponding to the absence of nuclear modification effects.
The measured cross sections are significantly lower than the IA predictions, indicating strong nuclear modifications in coherent \PgUa photoproduction.
To quantify the suppression, the ratio of the measured cross sections to the IA prediction, $S_{\PgUa}=\text{Data}/\text{IA}$, is shown in the lower panels of Fig.~\ref{fig:xSec}. The suppression factors are $S_{\PgUa}=0.25\pm0.06\stat\pm0.02\syst=0.25\pm0.07\,(\text{tot})$ at $\abs{y}<1.0$ and $S_{\PgUa}=0.32\pm0.09\stat\pm0.06\syst=0.32\pm0.11\,(\text{tot})$ at $1.0<\abs{y}<2.4$. The quoted uncertainties in $S_{\PgUa}$ are experimental only and do not include any uncertainty in the IA baseline. Notably, this suppression is only slightly larger than  those previously observed for the coherent \PGf photoproduction~\cite{CMS:2025lsm}, despite the probed $\mu^2$ differing by approximately two orders of magnitude. In both cases, the suppression factors are defined with respect to the same IA reference prescription.

A LO pQCD-based model emphasizing leading-twist contributions (referred to as LTA) incorporates nuclear shadowing effects arising from coherent multi-nucleon interference and provides both strong (SS) and weak shadowing (WS)  scenarios, with the SS scenario maximizing the contributions from three or more nucleons during the scattering process~\cite{Guzey:2016piu}. While both scenarios tend to overpredict the data, they show better agreement at $1.0<\abs{y}<2.4$.
The \STARLIGHT~\cite{Klein:2016yzr} model combines the vector meson dominance picture~\cite{Bauer:1977iq} with the classical Glauber formalism~\cite{Gribov:1968jf}, considering only elastic interactions, to describe the multiple scatterings of the photon-fluctuated virtual vector meson with the nucleus. The \STARLIGHT model significantly overestimates the data.

Rather than treating nuclear shadowing dynamically, the NLO pQCD framework of Ref.~\cite{Eskola:2023oos} computes the cross sections directly using nuclear PDFs (nPDFs) from global fits, such as EPS09~\cite{Guzey:2016piu} and EPPS21~\cite{Eskola:2023oos}. For the latter, the corresponding generalized parton distributions (GPDs) are modeled via the Shuvaev transform~\cite{Shuvaev:1999fm,Shuvaev:1999ce} of nPDFs. The predictions are labeled as ``NLO pQCD + EPS09'' and ``NLO pQCD + EPPS21-nGPDs'' in Fig.~\ref{fig:xSec}.
To reduce uncertainties in the $\mu^2$ choice and the tendency of NLO calculations to underpredict $\PGg\Pp$ photoproduction, Ref.~\cite{Eskola:2023oos} also provides an ``NLO pQCD + Data-Driven ($\PGg\Pp$)'' prediction, in which the nuclear gluon suppression factor is computed in NLO pQCD, while the \PgUa cross section in $\PGg\Pp$ interactions is constrained by HERA and LHC data~\cite{H1:2000kis,ZEUS:1998cdr,ZEUS:2009asc,LHCb:2015wlx,CMS:2018bbk}.
The NLO pQCD + EPS09 and NLO pQCD + Data-Driven ($\PGg\Pp$) predictions tend to overestimate the data, whereas the NLO pQCD + EPPS21-nGPDs result is consistent, although the large associated nPDF uncertainties limit conclusions based on central values.

In symmetric UPCs, a \PgUa meson measured at a given $y$ receives contributions from two photon energies of $(M_{\PgUa}/2)\re^{\mp y}$. The rapidity sign is defined by the photon-emitting nucleus in the laboratory frame, which is experimentally indistinguishable since either nucleus can act as the photon emitter or the target~\cite{Guzey:2013jaa,Guzey:2016piu,CMS:2023snh}.
Consequently, the measured $\rd\sigma/\rd y$ corresponds to the sum of two contributions at $x(\pm)=(M_{\PgUa}/\sqrtsNN)\re^{\pm y}$ and $\WgN(\pm)=\sqrt{\smash[b]{\sqrtsNN M_{\PgUa}\re^{\mp y}}}$.
At midrapidity ($\abs{y}<0.2$), the two photon energies are almost equal, allowing the averaged \PgUa production cross section per $\PGg\PbIon$ interaction to be obtained directly as
\begin{equation}
\sigma(\PGg\PbIon\to\PgUa\PbIon)=\frac{1}{2\nGama}\frac{\rd\sigma_{\PgUa}}{\rd y},
\label{eq:Wdependece}
\end{equation}
where \nGama is the photon flux~\cite{Klein:2016yzr}. The extracted $\sigma(\PGg\PbIon\to\PgUa\PbIon)$ corresponds to $\WgN\approx200\GeV$ and $x\approx10^{-3}$.
Uncertainties in the determination of \nGama, evaluated by varying the \PbIon nuclear radius and skin thickness within their uncertainties~\cite{CMS:2023snh,CMS:2025oxg,dEnterria:2020dwq}, amount to approximately 5\%.

\begin{figure}
    \centering
    \includegraphics[width=\cmsFigWidth]{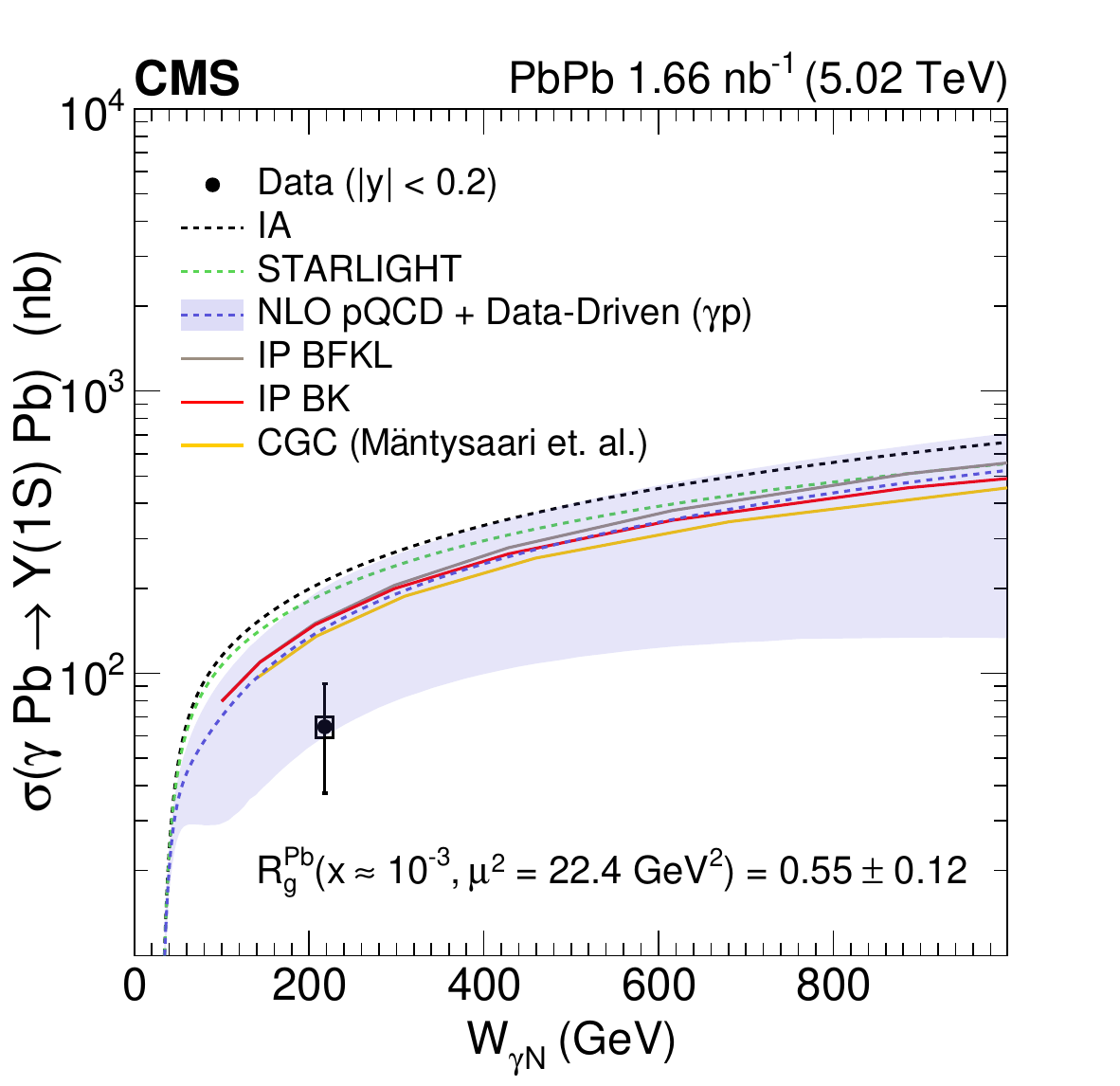}
    \caption{
        Coherent \PgUa photoproduction cross section per $\PGg\PbIon$ interaction as a function of \WgN. The vertical error bars and boxes represent statistical and systematic uncertainties, respectively.
        The data are compared with calculations based on the impulse approximation (IA)~\cite{Guzey:2013jaa,Klein:2016yzr}, \STARLIGHT~\cite{Klein:2016yzr}, the NLO pQCD + Data-Driven ($\PGg\Pp$) approach~\cite{Eskola:2023oos}, and several models (named ``IP BFKL'', ``IP BK'' and ``CGC'') within the CGC framework~\cite{Mantysaari:2022sux,Mantysaari:2023xcu,Penttala:2024hvp}. The quoted uncertainty in $\RgPb(x,\mu^2)$ represents the experimental uncertainty in $\sigma(\gamma\PbIon\to\PgUa\PbIon)$, including the contribution from the photon flux, and does not include any uncertainty in the IA baseline.
    }
    \label{fig:Wdependence}
\end{figure}

Nuclear effects on the gluon distribution in \PbIon are quantified using the nuclear gluon suppression factor $\RgPb(x,\mu^2)$, defined in Ref.~\cite{Guzey:2013xba} as $\sqrt{\smash[b]{\sigma^{\text{data}}/\sigma^{\text{IA}}}}$.
This definition reflects the approximate LO relation between the coherent photoproduction cross section and the square of the gluon distribution, which makes it possible to interpret \RgPb as a suppression factor at the level of the gluon density. Using $\sigma(\PGg\PbIon\to\PgUa\PbIon)$, $\RgPb(x,\mu^2)$ is evaluated at $x\approx10^{-3}$ and $\mu^2=22.4\GeV^{2}$, with $\mu=M_{\PgUa}/2$~\cite{ParticleDataGroup:2024cfk}, and the extracted value is $0.55\pm0.12\stat\pm0.02\syst=0.55\pm0.12\,(\text{tot})$, suggesting substantial nuclear gluon suppression. The quoted uncertainty represents the experimental uncertainty in $\sigma(\gamma\PbIon\to\PgUa\PbIon)$, including the contribution from the photon flux, and does not include any uncertainty in the IA baseline.

Figure~\ref{fig:Wdependence} shows the extracted $\RgPb(x,\mu^2)$ and $\sigma(\PGg\PbIon\to\PgUa\PbIon)$, with the latter compared as a function of \WgN to various theoretical predictions. Since \WgN determines the effective target momentum fraction $x$ probed in the interaction, the $\sigma(\PGg\PbIon\to\PgUa\PbIon)$ dependence on \WgN provides direct sensitivity to the energy evolution of nuclear gluon suppression. In addition to the IA, \STARLIGHT, and NLO pQCD + Data-Driven ($\PGg\Pp$) calculations, three predictions based on the color glass condensate (CGC) framework are also included~\cite{Mantysaari:2022sux,Mantysaari:2023xcu,Penttala:2024hvp}.
The CGC framework provides an effective description of high-density gluon fields at small $x$. The CGC prediction of Refs.~\cite{Mantysaari:2022sux,Mantysaari:2023xcu} is formulated in the dipole picture, in which the photon fluctuates into a quark-antiquark dipole that scatters from the target color field, with the small-$x$ evolution of the target encoded in Wilson lines evolved from a McLerran--Venugopalan initial condition using the Jalilian--Marian--Iancu--McLerran--Weigert--Leonidov--Kovner (JIMWLK) evolution. The ``IP BFKL'' prediction of Ref.~\cite{Penttala:2024hvp} uses an impact-parameter-dependent dipole framework in which the dipole amplitude evolves according to the linear Balitsky--Fadin--Kuraev--Lipatov (BFKL) equation~\cite{Balitsky:1978ic,Kuraev:1977fs,Fadin:1998py}, so the energy dependence is driven by linear small-$x$ evolution without the nonlinear saturation term.  The other prediction of Ref.~\cite{Penttala:2024hvp}  employs nonlinear Balitsky--Kovchegov (BK) evolution~\cite{Balitsky:1995ub,Kovchegov:1999ua}, which incorporates gluon recombination in addition to gluon splitting and dynamically generates saturation. This prediction is labeled ``IP BK'', and the parameters of its starting energy scale are fixed by fitting the calculation to HERA data. The similarity between the IP BFKL and IP BK predictions reflects the relatively large $\mu^2$ of coherent \PgUa photoproduction, for which nonlinear saturation effects are expected to be modest. Both predictions are also comparable to the NLO pQCD results and tend to overestimate the measured cross section.

In summary, the first measurement of coherent \PgUa meson photoproduction off a heavy nucleus is presented. The analysis is performed by the CMS experiment using ultraperipheral lead-lead collisions at a nucleon-nucleon center-of-mass energy of 5.02\TeV.
The  cross section is measured in two intervals of the \PgUa meson rapidity within $\abs{y}<2.4$. In the $\abs{y}<1$ interval, the ratio of the measured photoproduction cross section to a baseline model that neglects nuclear effects is $S_{\PgUa}=0.25\pm0.06\stat\pm0.02\syst$, thereby demonstrating nuclear suppression in this process. The nuclear gluon distribution is probed at  a nucleon momentum fraction of order $x{\sim}10^{-3}$ and at the unprecedented scale $\mu^2 = 22.4\GeV^2$ for coherent vector-meson photoproduction.  The \PgUa production cross section per photon-lead interaction yields a nuclear gluon suppression factor of $\RgPb(x\approx10^{-3},\mu^2=22.4\GeV^2)=0.55\pm0.12\stat\pm0.02\syst$ at $\abs{y}<0.2$, where the kinematics correspond to a photon-nucleon center-of-mass energy of ${\approx}200\GeV$. The measured $S_{\PgUa}$ remains remarkably close to the values observed in coherent \PGf photoproduction, despite the probed $\mu^2$ differing by approximately two orders of magnitude. Models incorporating nuclear shadowing or gluon saturation effects tend to underestimate the degree of suppression. Predictions based on next-to-leading-order perturbative quantum chromodynamics calculations  are consistent with the data only once the large uncertainties in the poorly constrained nuclear parton distribution functions are taken into account. 
Future higher precision measurements of the \PgU photoproduction at the LHC, as well as at planned facilities such as the Electron Ion Collider, will enable further exploration of mass-dependent nuclear suppression in vector-meson photoproduction and a broader \WgN reach.

\begin{acknowledgments}
We congratulate our colleagues in the CERN accelerator departments for the excellent performance of the LHC and thank the technical and administrative staffs at CERN and at other CMS institutes for their contributions to the success of the CMS effort. In addition, we gratefully acknowledge the computing centers and personnel of the Worldwide LHC Computing Grid and other centers for delivering so effectively the computing infrastructure essential to our analyses. Finally, we acknowledge the enduring support for the construction and operation of the LHC, the CMS detector, and the supporting computing infrastructure provided by the following funding agencies: SC (Armenia), BMBWF and FWF (Austria); FNRS and FWO (Belgium); CNPq, CAPES, FAPERJ, FAPERGS, and FAPESP (Brazil); MES and BNSF (Bulgaria); CERN; CAS, MoST, and NSFC (China); MINCIENCIAS (Colombia); MSES and CSF (Croatia); RIF (Cyprus); SENESCYT (Ecuador); ERC PRG and PSG, TARISTU24-TK10 and MoER TK202 (Estonia); Academy of Finland, MEC, and HIP (Finland); CEA and CNRS/IN2P3 (France); SRNSF (Georgia); BMFTR, DFG, and HGF (Germany); GSRI (Greece); NKFIH (Hungary); DAE and DST (India); IPM (Iran); SFI (Ireland); INFN (Italy); MSIT and NRF (Republic of Korea); MES (Latvia); LMTLT (Lithuania); MOE and UM (Malaysia); BUAP, CINVESTAV, CONACYT, LNS, SEP, and UASLP-FAI (Mexico); MOS (Montenegro); MBIE (New Zealand); PAEC (Pakistan); MSHE, NSC, and NAWA (Poland); FCT (Portugal); MESTD (Serbia); MICIU/AEI and PCTI (Spain); MOSTR (Sri Lanka); Swiss Funding Agencies (Switzerland); MST (Taipei); MHESI (Thailand); TUBITAK and TENMAK (T\"{u}rkiye); NASU (Ukraine); STFC (United Kingdom); DOE and NSF (USA).
\end{acknowledgments}
\bibliography{auto_generated}
\cleardoublepage \appendix\section{The CMS Collaboration \label{app:collab}}\begin{sloppypar}\hyphenpenalty=5000\widowpenalty=500\clubpenalty=5000
\cmsinstitute{Yerevan Physics Institute, Yerevan, Armenia}
{\tolerance=6000
A.~Belyaev\cmsorcid{0000-0003-1692-1173}, A.~Hayrapetyan, A.~Tumasyan\cmsAuthorMark{1}\cmsorcid{0009-0000-0684-6742}
\par}
\cmsinstitute{Institut f\"{u}r Hochenergiephysik, Vienna, Austria}
{\tolerance=6000
W.~Adam\cmsorcid{0000-0001-9099-4341}, J.W.~Andrejkovic, L.~Benato\cmsorcid{0000-0001-5135-7489}, T.~Bergauer\cmsorcid{0000-0002-5786-0293}, M.~Dragicevic\cmsorcid{0000-0003-1967-6783}, C.~Giordano\cmsorcid{0000-0001-6317-2481}, P.S.~Hussain\cmsorcid{0000-0002-4825-5278}, M.~Jeitler\cmsAuthorMark{2}\cmsorcid{0000-0002-5141-9560}, N.~Krammer\cmsorcid{0000-0002-0548-0985}, A.~Li\cmsorcid{0000-0002-4547-116X}, D.~Liko\cmsorcid{0000-0002-3380-473X}, M.~Matthewman, I.~Mikulec\cmsorcid{0000-0003-0385-2746}, J.~Schieck\cmsAuthorMark{2}\cmsorcid{0000-0002-1058-8093}, R.~Sch\"{o}fbeck\cmsAuthorMark{2}\cmsorcid{0000-0002-2332-8784}, D.~Schwarz\cmsorcid{0000-0002-3821-7331}, M.~Shooshtari\cmsorcid{0009-0004-8882-4887}, M.~Sonawane\cmsorcid{0000-0003-0510-7010}, W.~Waltenberger\cmsorcid{0000-0002-6215-7228}, C.-E.~Wulz\cmsAuthorMark{2}\cmsorcid{0000-0001-9226-5812}
% ...
\par}
\vskip\cmsinstskip
\dag:~Deceased\\
$^{1}$Also at Yerevan State University, Yerevan, Armenia\\
$^{2}$Also at TU Wien, Vienna, Austria\\
% ...
\end{sloppypar}
\end{document}